# System Synthesis for Networks of Programmable Blocks

Ryan Mannion, Harry Hsieh, Susan Cotterell, Frank Vahid*
Department of Computer Science and Engineering
University of California, Riverside
{rmannion, harry, susanc, vahid}@cs.ucr.edu
*Also with the Center for Embedded Computing Systems at UC Irvine

**Abstract**

*The advent of sensor networks presents untapped opportunities for synthesis. We examine the problem of synthesis of behavioral specifications into networks of programmable sensor blocks. The particular behavioral specification we consider is an intuitive user-created network diagram of sensor blocks, each block having a pre-defined combinational or sequential behavior. We synthesize this specification to a new network that utilizes a minimum number of programmable blocks in place of the pre-defined blocks, thus reducing network size and hence network cost and power. We focus on the main task of this synthesis problem, namely partitioning pre-defined blocks onto a minimum number of programmable blocks, introducing the efficient but effective PareDown decomposition algorithm for the task. We describe the synthesis and simulation tools we developed. We provide results showing excellent network size reductions through such synthesis, and significant speedups of our algorithm over exhaustive search while obtaining near-optimal results for 15 real network designs as well as nearly 10,000 randomly generated designs.*

## 1. Introduction

The availability of low cost, low power electronics has made ubiquitous monitoring and control realizable through the advent of sensor networks. Monitoring and controlling various aspects of a home, office, store, school, factory, or public spaces can result in improved personal comfort and safety, as well as public security. However, today's monitor/control systems are largely specialized, factory-programmed and configured to perform a specific application, such as home automation or intrusion detection. Though the applications of monitor/control systems are voluminous, high design cost has restricted all but the most commercially viable products from entering the market, resulting in an unfulfilled void of useful applications. For example, a homeowner may want notification of a garage door open at night or of a sleepwalking child; an office worker may want to know whether mail exists for him in the mailroom, a copy machine is free, or if conference rooms are in use. These applications are useful but lack the volume to justify producing dedicated products to address them, or results in high-priced dedicated products such that contracting a customized system would be cost prohibitive. A generalized set of building blocks that could be combined to perform a broad range of functions while maintaining ease of use would therefore serve a great need for non-programmer, non-engineer users to construct a variety of monitor/control systems.

Technological advances in miniaturization, wireless communication, and power efficiency have recently made feasible the development of low-cost, intelligent, and configurable networks of devices, known generally as sensor networks [2][11][17][19]. Returning to a previous example, instead of a dedicated garage open at night product, a user can connect a set of *inexpensive reusable* electronic blocks to build a system that accomplishes the same task. A garage open at night system would require a contact switch block, a light sensor block, a logic block, and an output block (e.g. buzzer or LED (light-emitting diode)) placed perhaps in a bedroom. A sleepwalk detector would utilize a motion sensor block, light sensor block, logic block and output block. A copy machine use detector might use just a motion sensor and output block. A conference room in-use detector might use motion and sound sensor blocks, logic blocks, and output blocks. Notice that the same blocks are usable in a variety of applications, enabling mass production and hence commercially viable low-cost blocks.

While many types of sensor networks utilize general-purpose programmable blocks [11][19], others have proposed using blocks with pre-defined functions and communication protocol [6][13][15][16][20]. Such pre-defined blocks enable unskilled people (in this case, people with no programming or hardware experience) to still build basic but useful systems, and may even reduce design time for skilled designers. However, while pre-defined blocks have ease of use advantages, they may result in more blocks in a network, resulting in larger network size, higher cost, and increased power consumption.

Therefore, we have developed capture, simulation and synthesis tools that enable a user to specify a network using pre-defined blocks (representing a behavioral description) and to simulate that network to verify correct behavior, and that automatically synthesize an optimized network using programmable blocks with automatically generated software.

We found a challenging part of the development of these tools to be the design of the algorithm for partitioning the behavioral description onto a minimum set of programmable blocks. Although we originally assumed the partitioning problem would be equivalent to standard partitioning problems or to existing technology mapping problems, we found that particular differences made impossible the use of existing algorithms – hence, we developed a new decomposition algorithm to solve the problem.

In the following section, we discuss background work, and provide justification for our work in this area. In Section 3, we discuss the specification, the simulator, and the code generation aspect of the system synthesis. Section 4 focuses on the partitioning aspect of the synthesis process and introduces our algorithm for partitioning. Section 5 provides the experimental results on real designs, as well as on large randomly generated designs. We conclude in Section 6.

## 2. Background

Much of the previous work in sensor networks relates to development of general-purpose sensor nodes [11][19]. Synthesis related works focus largely on analysis or exploration [1][10][18][21] of communication issues for networks having large numbers (thousands) of irregularly structured networked wireless nodes.

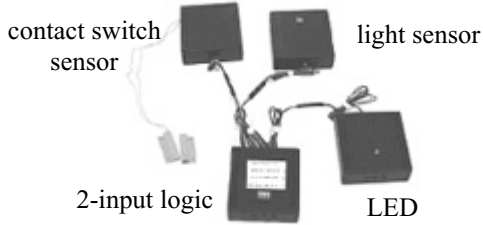

**Figure 1**: Garage-open-at-night system built using eBlock prototypes.

contact switch sensor
light sensor
2-input logic
LED

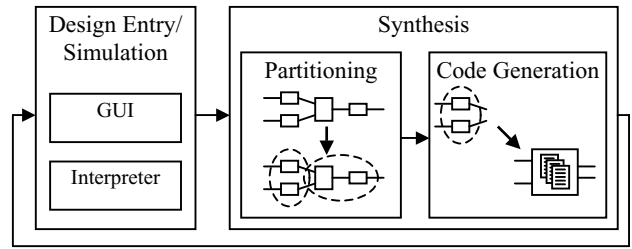

**Figure 2**: Design framework.

Design Entry/Simulation
GUI
Interpreter

Synthesis
Partitioning
Code Generation

Our focus is instead on smaller-scale networks whose nodes have particular functions and whose connectivity explicitly carries out a particular task [6][13][15][16][20], such as a garage open at night detector. In particular, we focus on eBlocks [6][7]. eBlocks feature four classes of "blocks," categorized by function, communicating with a standard protocol. Sensor blocks detect environmental stimuli, such as motion or the press of a button. Output blocks interact with the environment, such as emitting a beep or turning on an appliance via an electric relay. Though blocks normally connect directly to each other using wires, communication blocks allow for wireless communication or other methods (such as X10 [22]). Compute blocks perform a (typically pre-defined) function on inputs and generate output. A special type of compute block is a programmable block, which features a finite number of inputs and outputs and which can be programmed to implement custom functionality (including functionality of multiple pre-defined blocks). Pre-defined compute functions include combinational functions, such as a two or three input truth table, AND, OR, and NOT, and basic sequential functions, like a toggle, trip, pulse generate, and delay. Figure 1 shows a garage open at night eBlock system built from physical eBlocks.

We developed a simulator for all four classes of eBlocks and their behavioral interaction. The synthesis procedure, including that of design partitioning, can then reduce the total number of eBlocks by combining the predefined compute blocks into a smaller number of programmable compute blocks.

Design partitioning is a critical aspect for successful system level synthesis. During partitioning, we seek to replace clusters of eBlocks with programmable eBlocks and minimize the total number of eBlocks. This partitioning problem cannot be solved using approaches developed for the classic bin-packing problem or the knapsack problem, as we must mind at least two constraints when "packing" a programmable block: the number of inputs used and the number of outputs used. Solutions to the two-dimensional bin-packing problem, commonly manifested in the *cutting stock problem* [3], are also not applicable to our problem. The input and output constraints of the programmable blocks are mutually independent, the number of inputs used in a programmable block has no effect on the number of outputs available, and vice versa, and thus our problem cannot be restated as a cutting stock problem.

Our partitioning problem is similar to the problem of synthesizing behavioral specifications to FPGAs (Field Programmable Gate Arrays). Specifically, our problem is closest to that of DAG (Directed Acyclic Graph) covering. DAG covering, as it relates to our problem, has been shown to be NP-hard [12]. However, algorithms developed for finding solutions to DAG covering, particularly algorithms that map Boolean networks to LUTs (look-up tables), are not applicable to our problem for a number of reasons. First, we do not require that all nodes in our network graph be covered (i.e. not all blocks need to be replaced by a programmable block). In fact, circumstances in which a partition contains a single node are undesirable, as there is no net reduction in the size of the design with a one-to-one mapping of a pre-defined block to a programmable block. This relaxation of the problem increases the search space significantly as we must now consider partitions in which one or more nodes are not covered. Second, we define our optimal solution as a cover that covers the most number of blocks with the fewest number of partitions, analogous to the goal of minimum-area DAG covering. However, many DAG covering algorithms focus on finding minimum-delay solutions or approximations [4][5][14]. Third, many existing DAG covering algorithms permit replication of nodes in each LUT/programmable block [9], but duplicating code in the programmable blocks runs contrary to the need to minimize power usage in sensor networks.

Thus, we contend that a partitioning algorithm targeted to the problem of partitioning pre-defined compute blocks to a minimum number of programmable compute blocks with limited input/output solves a new and useful problem.

## 3. Design Framework
We have written a number of tools that enable the specification, simulation, partitioning, and code generation of eBlock systems. Figure 2 illustrates our tool chain.

### 3.1 Design Entry/Simulation
We developed a Java-based graphical user interface (GUI) and simulator for the design entry and simulation of an eBlock system, illustrated in Figure 3. A user can drag a block from a library of blocks, situated on the right edge of the simulator, to the workspace and connect the blocks together by drawing lines between circular representations of their input/outputs. The simulator incorporates eBlocks that sense or interact with their environment with an accompanying visual representation of their environmental stimuli/interaction (e.g. clicking on a light sensor will toggle a light bulb icon on and off).

The simulator contains behavioral representations of all available eBlocks. It also understands the underlying communication protocol and all the restrictions on the model of computation. The simulation is behaviorally correct and obeys general high-level timing, but no detailed timing characteristics can be inferred. Given that all communication between blocks is done serially using packets and hence globally asynchronous, and because the blocks deal with human-scale events rather than fast timing, such lack of timing detail is generally not a problem.

### 3.2 eBlock Partitioning
After completing a design in the simulator, the user can instruct the framework to minimize the number of computation eBlocks needed by implementing them with a minimum number of programmable eBlocks. The simulator passes the design into our partitioning tool, which produces a list of partitions based on our partitioning algorithm. The partitioning tool subsequently passes each partition to the code generation tool, which translates the interaction between the partition's inputs and component blocks into sequential code that can run on a programmable block.

**Figure 3**: eBlock graphical capture and simulator tool.

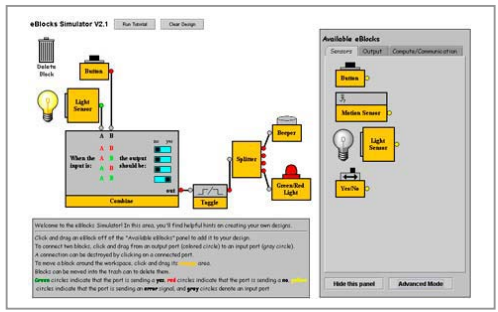

Using the eBlock Design Framework, we can utilize different partitioning algorithms for different designs. Detailed discussion of the partitioning problem and the algorithms used will be presented in Section 4.

### 3.3 Code Generation

Each partition generated by the partitioning tool utilizes the code generation tool. The tool assigns each block within a partition a level, defined as the maximum distance between the block and any sensor block (analogous to the primary input-based level definition in circuit partitioning). The code generation tool assigns levels by tracing the paths in the network, beginning with sensor blocks, and marking each block visited with an appropriate level. Blocks visited multiple times retain the greatest level value assigned to them. eBlock networks do not contain loops, so the levels for eBlocks in a network are well-defined.

The simulator maintains the behavior of each block, defined in a Java-like language that is automatically transformed to a syntax tree. The code generator attaches the syntax trees of every block in the partition so that the tool can evaluate trees in succession. Syntax trees are ordered in non-decreasing order and determined by the level of each block. The ordering of the syntax trees ensures that the tool does not evaluate a block's tree before any of its input blocks have produced output. As the tool merges each block's tree into the new programmable block's tree, the tool changes tree nodes that access a block's input or output into a variable access, where appropriate. Communication between two blocks in a partition will occur internally in a programmable block via variables. In the event that two or more blocks share variable names in their internal behavior code, the conflict is resolved through variable renaming.

The newly merged syntax tree specifies the behavior of the programmable block that will replace the partition. The simulator's interpreter evaluates the tree in the same manner as a non-programmable block. A user can select a programmable block and instruct the simulator to translate the syntax tree into C code for downloading and use in a physical block.

Since we have implemented physical programmable eBlock system prototypes, we were able to compile and download the generated code and use the output of the synthesis tool chain in real-world systems. The programmable eBlock prototype utilizes a Microchip PIC16F628 microcontroller (www.microchip.com) with 2 Kbytes of program memory and is available for under $2. Given that amount of program memory, the small size of each program describing a pre-defined block's function, and the scale of real eBlock systems, we make the practical assumption that a programmable block's program size constraint will not be violated by any partition. Making such violation even less likely is the fact that the block-partitioning problem is actually input/output limited, not size limited. However, even if a programmable block's memory size were exceeded, we could easily utilize a PIC with a larger memory, or extend our algorithm with size constraints.

### 4. Partitioning

As mentioned in Section 2, the general aim of the partitioning phase is to minimize the total number of blocks by replacing the greatest number of pre-defined compute blocks with fewer programmable blocks. Pre-defined compute blocks have identical internal components and thus have equal cost. A programmable compute block has slightly higher cost due to the programmability hardware, but less cost than two pre-defined compute blocks.

A more precise statement of the problem is as follows. We represent an eBlock system as a directed acyclic graph $G = (V, E)$ where $V$ is the set of nodes (blocks) in the graph and $E$ is the set of edges (connections) between the nodes. We define sensor eBlocks as primary inputs, and output eBlocks as primary outputs. The objective of the partitioning phase is to find a set of subgraphs of non-primary input and non-primary output nodes (hereafter referred to as *inner* nodes) of $G$ such that 1) each subgraph has at most $i$ inputs and $o$ outputs, where $i$ and $o$ correspond to the number of inputs and outputs available in a programmable block, 2) each subgraph must be replaceable by a programmable block that can provide equivalent functionality, and 3) the number of inner blocks after replacement is minimized. We assume a subgraph containing only a single node to be invalid, as it is undesirable to replace a pre-defined compute block with a programmable compute block due to slightly higher cost of the latter.

### 4.1 Exhaustive Search

Initially, we implemented an exhaustive search algorithm to see whether an exhaustive approach was reasonable with respect to computation time, and to provide optimum results for small problem instances. The search space consists of every combination of $n$ blocks into $n$ programmable blocks (a combination need not use every block). We utilized a simple pruning mechanism in which all "empty" programmable blocks in a combination are indistinguishable, and portions of the search tree removed accordingly. For example, if at a given point in the search tree we are able to assign a block to one of three empty programmable blocks, we only consider one such branch.

The run time of partitioning by exhaustive search naturally increased exponentially. With designs featuring fourteen inner blocks, the search did not conclude after four hours. Such run time of exhaustive search is not acceptable for designs with as few as eleven inner blocks, where a user must wait for about one minute. An exhaustive search is therefore overly prohibitive as eBlocks designs with eleven or more inner blocks are common, and thus a different approach is needed.

**Figure 4**: Decomposition method pseudocode.

```
PareDown heuristic
  blocks ← list of inner blocks
  partitions ← empty list
  while blocks contains elements
    partition ← blocks
    while partition contains elements
      if partition fits in a programmable block then
        if partition contains more than one block then
          add partition to partitions
        else if partition contains zero blocks
          return partitions
        end if
        remove elements in partition from blocks
      else
        compute ranks for border blocks in partition
        remove border block from partition with the least rank
      end if
    end while
  end while
  return partitions
  (partitions contains a list of partitions)
```

## 4.2 Decomposition Method – The "PareDown" Heuristic

We first implemented a heuristic that clusters nodes into subgraphs through *aggregation*. From a list of inner nodes connected to a primary input, the aggregation method repeatedly selects a node that fits within a programmable block as a partition. Though the aggregation method concludes quickly for cases in which the exhaustive search is impractical, the aggregation method is not capable of taking advantage of convergence and thus we found it often produced non-optimal results.

We sought to avoid the pitfalls of the aggregation method's lack of look-ahead capabilities, by utilizing a *decomposition method* heuristic we call PareDown, to detect convergence but not be constrained by a particular "depth" at which our heuristic looks ahead. Thus, the PareDown approach, shown in Figure 4, begins by selecting all internal blocks of a design as a candidate partition, and then removes blocks from the partition until input and output constraints are met, hence the moniker "decomposition method." If the algorithm finds a valid partition, the algorithm repeats on the remaining blocks and continues until no new partitions remain.

The choice of which block to remove from an invalid candidate partition is determined by computing a rank for all *border* blocks in the candidate partition and removing the block with the lowest rank. We define a border block as a block in which every output or every input connects to a block outside of the candidate partition. The block's rank is defined as the net increase or decrease in the combined indegree and outdegree of a candidate partition if that block is removed from the candidate partition. For blocks that have the same rank, removal priority is given to the following criteria, in order: 1) The block with the greatest indegree 2) The block with the greatest outdegree 3) The block with the highest level.

The decomposition method benefits from fast run times and produces results well beyond the practical range of the exhaustive search. A worst-case design would feature inner blocks that can fit into a partition by themselves but cannot form a valid partition when combined with any other block or blocks. A design of this form containing *n* inner blocks would therefore require *n* iterations through the algorithm's main loop. In the first iteration, the PareDown heuristic would check for a valid partition *n* times before eventually isolating a single block as a partition (an invalid result). Subsequent iterations check for valid partitions *n-1* times and so forth. Accordingly, the total iterations through the decomposition method is *n*(n+1)/2*, or the sum of 1 to *n*. This total yields a time complexity of $O(n^2)$.

### 4.2.1 Example

As an example, we will demonstrate the PareDown heuristic partitioning of an example eBlock design, the Podium Timer 3 design from [8]. The podium timer utilizes a set of eBlocks to build a system that can notify a person that his/her talk time is nearing the end, and that time has ended. The design consists of some buttons, output LEDs, and several internal pre-defined compute nodes, including combinational and sequential nodes.

We begin with the DAG representation of the design, shown in Figure 5(a). For ease of reference, we number each block. Figure 5(a) shows the initial selection by the PareDown heuristic of every interior block as a candidate partition, indicated by the shaded area, and the rank values for the border nodes 2, 8, and 9. We assume that the programmable block available features two inputs and two outputs.

Since the shaded partition in Figure 5(a) requires three outputs, the partition is invalid. The PareDown heuristic removes node 9, the node with the least rank value, as shown in Figure 5(b). The partition remains invalid and nodes 2 and 8 are considered for removal, being the border nodes (node 6 and 7 are not border nodes since one or more of their outputs connects to a block inside the candidate partition). The two border nodes have the same rank but node 8 has a greater indegree and is removed, as shown in Figure 5(c).

With a requirement of four outputs, the partition shown in Figure 5(c) is invalid. Subsequent steps remove nodes 7 and 6 from the partition, and the partition becomes valid, as indicated by the solid outline in Figure 5(d).

The algorithm is repeated on the remaining node(s): 6, 7, 8, and 9. The PareDown heuristic removes node 7 from the new candidate partition, and the new partition becomes valid. Again, the heuristic runs on the un-partitioned node(s), which consists only of node 7. Though the partition fits in a programmable block,

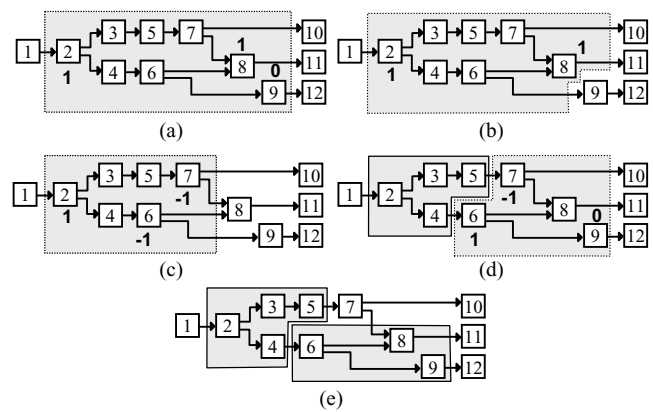

**Figure 5**: Decomposition method example.

Table 1: Results for exhaustive search and PareDown decomposition using design library.

| Inner Blocks (Original) | Design Name | Averages for Exhaustive Search | | | Averages for "PareDown" Decomposition | | | | |
|---|---|---|---|---|---|---|---|---|---|
| | | Inner Blocks (Total) | Inner Blocks (Prog.) | Time | Inner Blocks (Total) | Inner Blocks (Prog.) | Time | Block Overhead | % Overhead |
| 2 | Ignition Illuminator | 1 | 1 | <1ms | 1 | 1 | <1ms | 0 | **0 %** |
| 2 | Night Lamp Controller | 1 | 1 | <1ms | 1 | 1 | <1ms | 0 | **0 %** |
| 2 | Entry Gate Detector | 1 | 1 | <1ms | 1 | 1 | <1ms | 0 | **0 %** |
| 2 | Carpool Alert | 1 | 1 | <1ms | 1 | 1 | <1ms | 0 | **0 %** |
| 3 | Cafeteria Food Alert | 1 | 1 | <1ms | 1 | 1 | <1ms | 0 | **0 %** |
| 3 | Podium Timer 2 | 1 | 1 | <1ms | 1 | 1 | <1ms | 0 | **0 %** |
| 3 | Any Window Open Alarm | 3 | 0 | <1ms | 3 | 0 | <1ms | 0 | **0 %** |
| 3 | Two Button Light | 3 | 1 | <1ms | 3 | 1 | <1ms | 0 | **0 %** |
| 5 | Doorbell Extender 1 | 5 | 0 | <1ms | 5 | 0 | <1ms | 0 | **0 %** |
| 6 | Doorbell Extender 2 | 6 | 0 | 9ms | 6 | 0 | <1ms | 0 | **0 %** |
| 8 | Podium Timer 3 | 3 | 3 | 125ms | 3 | 2 | <1ms | 0 | **0 %** |
| 10 | Noise At Night Detector | 6 | 4 | 4.79 s | 6 | 4 | <1ms | 0 | **0 %** |
| 19 | Two-Zone Security | -- | -- | -- | 10 | 3 | <1ms | -- | -- |
| 19 | Motion on Property Alert | -- | -- | -- | 19 | 0 | <1ms | -- | -- |
| 23 | Timed Passage | -- | -- | -- | 14 | 5 | <1ms | -- | -- |

*-- = no data available*

the partition is invalid for containing only a single block. The heuristic halts, returning the result visualized in Figure 5(e).

Thus, the heuristic reduces the internal compute nodes from the initial user-defined 8 nodes to only 3 programmable nodes.

## 5. Experiments
### 5.1 Setup
We executed the decomposition and exhaustive search algorithms on a variety of designs of varying depths (maximum block level) and size on a 2GHz AMD Athlon XP PC. The designs included 15 actual eBlock systems appearing at [8], designed independently of our work for purposes other than synthesis. Some of those systems were intentionally designed to be small by avoiding multiple identical sensors whenever possible to maximize user-comprehension of the systems, whereas eBlock systems can easily involve several dozen nodes. Thus, we also developed a randomized eBlock system generator able to generate eBlock networks of varying sizes. We implemented both exploration algorithms, exhaustive and PareDown, in Java within the framework provided by the eBlocks simulator, with the two implementations each representing less than 100 lines of code, including the code to extract experimental data. We assumed the existence of a programmable block having two inputs and two outputs.

Table 1 summarizes the data collected during the experiments for the library-based designs. The *Inner Blocks (Original)* column indicates the number of inner blocks corresponding to the design specified by the *Design Name* column. The *Inner Blocks (Total)* column indicates the number of inner blocks in a given design after partitioning by either the exhaustive search or PareDown heuristic algorithm. The *Inner Blocks (Prog.)* column specifies, of the number of inner blocks after partitioning listed in the *Inner Blocks (Total)* column, the number of programmable blocks in the partitioned design, which is essentially the number of partitions found. The *Time* column specifies the execution time for either of the algorithm utilized. The *Block Overhead* column indicates the number of additional inner blocks required after partitioning by the heuristic algorithm compared to optimal results of the exhaustive search algorithm. The *% Overhead* indicates the percent increase in the final design when the heuristic is utilized compared to the optimal results of the exhaustive search algorithm.

In addition to the library of designs used in our experiments, we ran the exhaustive search and heuristic algorithms on a sample of randomly generated designs. Table 2 shows the corresponding results. The random designs' inner block count varied between three and forty-five inner blocks as illustrated in the *Inner Blocks (Original)* column. The *Number of Designs* column specifies the number of designs considered for a given inner block counts.

### 5.2 Runtime Performance
The decomposition algorithm achieves significant speedup over the exhaustive search throughout the entire range of inner block values. Though exhaustive search maintains a reasonable runtime up to ten inner blocks, adding more inner blocks has a dire effect on the exhaustive search's computation time. Inner block sizes of thirteen or more yield unacceptable wait times for a user.

Beyond thirteen inner blocks, the decomposition method continues to process large designs in a reasonable amount of time and is therefore a good replacement for the exhaustive search. Though not displayed in Table 2, the decomposition method produced a result for a design with 465 inner nodes in 80 seconds. We note that it is highly unlikely that a real-world eBlocks design would feature 465 inner nodes. Thus, our method should be sufficiently fast for all practical eBlock designs.

### 5.3 Optimality
The decomposition method achieves excellent results that are optimal or within 15% of an optimal solution for every inner block size in which data was available from the exhaustive search. As indicated in Table 1 and Table 2, the solution generated by the decomposition method for the displayed range of inner block sizes is within one block of an optimal solution, on the average.

## 6. Conclusions and Further Work
We have demonstrated a working system for synthesizing behavioral models of eBlock networks into code and connectivity for physical eBlocks. We focused on the partitioning aspect of synthesis and introduced the PareDown heuristic. We have shown that PareDown is an effective method for rapidly finding optimal or near-optimal solutions to eBlock partitioning for both real-world designs and generated ones.

We plan to extend the PareDown heuristic to consider multiple types of programmable blocks (having different number of inputs and outputs) and varying compute block costs, and to

Table 2: Results for exhaustive search and PareDown decomposition using randomly generated designs.

| Inner Blocks (Original) | Number of Designs | Averages for Exhaustive Search | | | Averages for "PareDown" Decomposition | | | | |
|---|---|---|---|---|---|---|---|---|---|
| | | Inner Blocks (Total) | Inner Blocks (Prog.) | Time | Inner Blocks (Total) | Inner Blocks (Prog.) | Time | Block Overhead | % Overhead |
| 3 | 1531 | 1.83 | 0.81 | <1ms | 1.87 | 0.79 | <1ms | 0.04 | 2 % |
| 4 | 982 | 2.24 | 1.22 | <1ms | 2.33 | 1.10 | <1ms | 0.09 | 4 % |
| 5 | 542 | 2.51 | 1.52 | 1.33ms | 2.62 | 1.32 | <1ms | 0.11 | 4 % |
| 6 | 432 | 3.08 | 1.74 | 6.56ms | 3.36 | 1.49 | <1ms | 0.28 | 9 % |
| 7 | 447 | 3.77 | 2.00 | 25.52ms | 4.09 | 1.73 | <1ms | 0.32 | 8 % |
| 8 | 350 | 4.11 | 2.32 | 122.97ms | 4.56 | 1.93 | <1ms | 0.45 | 11 % |
| 9 | 340 | 4.67 | 2.60 | 719.90ms | 5.24 | 2.17 | <1ms | 0.57 | 12 % |
| 10 | 199 | 5.04 | 2.93 | 4.53s | 5.76 | 2.45 | <1ms | 0.69 | 14 % |
| 11 | 170 | 5.47 | 3.20 | 31.77s | 6.29 | 2.59 | <1ms | 0.82 | 15 % |
| 12 | 31 | 4.58 | 3.23 | 3.67min | 4.87 | 2.58 | <1ms | 0.29 | 6 % |
| 13 | 6 | 6.84 | 3.17 | 29.93min | 7.83 | 2.83 | <1ms | 0.99 | 14 % |
| 14 | 1311 | -- | -- | -- | 8.11 | 3.05 | <1ms | -- | -- |
| 15 | 1184 | -- | -- | -- | 8.67 | 3.32 | <1ms | -- | -- |
| 20 | 928 | -- | -- | -- | 11.09 | 4.70 | <1ms | -- | -- |
| 25 | 691 | -- | -- | -- | 13.93 | 5.97 | 1.86ms | -- | -- |
| 35 | 354 | -- | -- | -- | 19.63 | 8.26 | 4.82ms | -- | -- |
| 45 | 165 | -- | -- | -- | 25.43 | 10.62 | 13.28ms | -- | -- |

*-- = no data available*

extend our methods to map to an existing underlying network of sensor nodes.

## 7. Acknowledgements

This work was supported by the National Science Foundation (CCR-0311026) and by a Department of Education GAANN fellowship, and aided by donations from Microchip Technology Corporation.